\documentclass[10pt,conference]{IEEEtran}
\IEEEoverridecommandlockouts

\usepackage{cite}
\usepackage{amsmath,amssymb,amsfonts}
\usepackage{algorithmic}
\usepackage{graphicx}
\usepackage{textcomp}
\usepackage{multirow}
\usepackage{booktabs}
\usepackage{array}
\usepackage{hyperref}

\usepackage{xcolor}
\def\BibTeX{{\rm B\kern-.05em{\sc i\kern-.025em b}\kern-.08em
    T\kern-.1667em\lower.7ex\hbox{E}\kern-.125emX}}
\begin{document}

\title{PyExamine: A Comprehensive, Un-Opinionated Smell Detection Tool for Python}

\author{\IEEEauthorblockN{Karthik Shivashankar}
\IEEEauthorblockA{\textit{Department of Informatics},
\textit{University of Oslo, Norway}\\
karths@ifi.uio.no}
\and
\IEEEauthorblockN{Antonio Martini}
\IEEEauthorblockA{\textit{Department of Informatics},
\textit{University of Oslo, Norway}\\
antonima@ifi.uio.no}}

\maketitle

\begin{abstract}
The growth of Python adoption across diverse domains has led to increasingly complex codebases, presenting  challenges in maintaining code quality. While numerous tools attempt to address these challenges, they often fall short in providing comprehensive analysis capabilities or fail to consider Python-specific contexts. PyExamine addresses these critical limitations through an approach to code smell detection that operates across multiple levels of analysis.

PyExamine architecture enables detailed examination of code quality through three distinct but interconnected layers: architectural patterns, structural relationships, and code-level implementations. This approach allows for the detection and analysis of 49 distinct metrics, providing developers with an understanding of their codebase's health. The metrics span across all levels of code organization, from high-level architectural concerns to granular implementation details.

Through evaluation on 7 diverse projects, PyExamine achieved detection accuracy rates: 91.4\% for code-level smells, 89.3\% for structural smells, and 80.6\% for architectural smells. These results were further validated through extensive user feedback and expert evaluations, confirming PyExamine's capability to identify potential issues across all levels of code organization with high  recall accuracy.

In additional to this, we have also used PyExamine to analysis the prevalence of different type of smells, across 183 diverse Python projects ranging from small utilities to large-scale enterprise applications. 

PyExamine's distinctive combination of comprehensive analysis, Python-specific detection, and high customizability makes it a valuable asset for both individual developers and large teams seeking to enhance their code quality practices. 

\end{abstract}

\begin{IEEEkeywords}
Python, Code Smells, Anti-patterns
\end{IEEEkeywords}

\section{Introduction}

Code smells are indicators of potential problems in software design and implementation \cite{fowler2018refactoring}. While not bugs themselves, these smells often signal design weaknesses that can impede development and increase the risk of future failures. As Python's popularity grows across data science, web development, and automation domains, maintaining large Python codebases has become increasingly challenging \cite{chen2016detecting} \cite{sandouka2023python}.

Current Python code analysis tools exhibit significant limitations in their detection capabilities. These tools, such as Pylint and Flake8 \cite{flake8} \cite{pylint}, primarily focus on code-level issues, lacking the ability to detect system-wide architectural problems like cyclic dependencies, god components, and scattered functionality \cite{vavrova2017does, giordano2023understanding} .

This limited architectural analysis is a major drawback in existing solutions. They often lack depth in their analysis capabilities, particularly when dealing with architectural concerns and complex structural patterns. Additionally, most tools offer limited customization options, making it difficult to adapt to project-specific requirements and coding standards. Perhaps most critically, these tools frequently struggle to integrate different levels of analysis effectively, leaving significant gaps in their detection capabilities\cite{garcia2009toward, mumtaz2021systematic}.

While tools like Prospector and Radon \cite{prospector, radon} provide some structural metrics, they fall short in effectively identifying complex inheritance hierarchies, poor interface cohesion, and problematic design patterns \cite{lanza2007object}. 

This insufficient structural detection leaves many potential issues unaddressed. Moreover, most existing tools lack proper consideration of Python's unique features and idioms, leading to unreliable detection and false positives \cite{vavrova2017does, boutaib2021code}.

Another significant limitation is the integration gap present in current solutions. These tools often operate in isolation, failing to connect code-level issues with their broader structural and architectural implications \cite{sas2022relation, arcelli2023impact}. This disconnection hampers a holistic understanding of the codebase and its potential problems.

The absence of comprehensive smell detection capabilities in Python projects has several important implications. First, without proper architectural smell detection, projects accumulate technical debt through unidentified cyclic dependencies and problematic component interactions \cite{martini2018identifying, suryanarayana2014refactoring}. This architectural debt can significantly impact the long-term maintainability and scalability of the software.

Second, the limited structural analysis capabilities lead to undetected design issues that compound over time, making maintenance increasingly difficult \cite{fontana2017arcan, sas2022evolution}. As these structural problems grow, they can severely hinder the evolution and adaptability of the codebase.

Lastly, the lack of multi-level analysis makes it challenging to understand how code-level issues impact broader architectural concerns \cite{palomba2013detecting, fontana2015automatic}. This integration challenge prevents developers from gaining a detailed view of their project's health and potential areas for improvement. Addressing these limitations is crucial for developing more effective and holistic Python code analysis tools.

To address these limitations, we introduce PyExamine, a comprehensive code smell detection tool specifically designed for Python projects. PyExamine employs a multi-level analysis approach that examines code at three distinct levels:

\begin{enumerate}
\item Architectural smells: The tool examines system-wide design patterns, dependencies, and package hierarchies using advanced graph-based algorithms. This includes detection of cyclic dependencies, analysis of component coupling, and evaluation of architectural layering violations. The architectural analysis employs metrics to identify issues that could impact system maintainability and scalability \cite{garcia2009toward, arcelli2023impact, mumtaz2021systematic}.
\item Structural smells: At this level, PyExamine evaluates class and module design, interface cohesion, and inheritance structures using established metrics such as LCOM (Lack of Cohesion of Methods), CBO (Coupling Between Objects), and DIT (Depth of Inheritance Tree). The tool also analyzes package-level organization and component interactions to identify structural weaknesses \cite{gupta2022prioritizing, gupta2023severity, mo2019architecture, lanza2007object, sahraoui2000can}.
\item Code-level smells: The most granular level of analysis identifies implementation issues through detailed static analysis of functions, methods, and code blocks. This includes detection of long methods, complex conditional logic, duplicate code, and other common code-level smells. The analysis employs sophisticated pattern matching and metric calculations to ensure accurate detection \cite{cardozo2023prevalence, foidl2022data, martin2009clean, palomba2013detecting, boutaib2021code}.
\end{enumerate}

PyExamine distinguishes itself through several core capabilities. The tool implements a multi-tiered analysis framework that ensures comprehensive smell detection across different abstraction levels. Its highly customizable configuration system enables developers to adapt the analysis to their project-specific requirements.

PyExamine incorporates an extensive catalog of Python-specific code smells, focusing on issues arising from Python's unique features and idioms. The tool covers 49 distinct metrics across three levels of analysis: 24 code-level metrics, 6 architectural metrics, and 19 structural metrics. Additionally, PyExamine provides a sophisticated reporting system that delivers clear, actionable insights for improvement.

Our evaluation across 7 diverse projects demonstrates PyExamine's effectiveness, achieving  Recall accuracy of 91.4\% for code-level smells and 89.3\% for structural smells and 80.6\% for Architectural smells. While recent research shows promise in using machine learning for smell detection \cite{sandouka2023python}, these approaches currently exhibit limitations in functionality and higher false-positive rates compared to rule-based methods.

Our study addresses four fundamental questions regarding smells detection:

\begin{enumerate}
    \item How accurate is PyExamine in detecting code smells?
    \item How accurate is PyExamine in detecting structural smells?
    \item How accurate is PyExamine in detecting architectural smells?
    \item What is the prevalence and distribution of different smell types in Python projects?
\end{enumerate}

Through this research, we aim to advance software quality assurance practices and provide developers with an effective tool for maintaining high-quality Python codebases \cite{chen2016detecting} \cite{sharma2018survey}. PyExamine represents a  step forward in automated code quality assessment, offering a comprehensive solution to the challenges of maintaining large-scale Python projects.

\section{Background and Related Work}
\subsection{Code, Architectural and Structural Smells}
Smells are symptoms in the source code that indicate potential design or implementation problems. While not bugs or errors in themselves, code smells often correlate with deeper issues in the software's design or architecture. They can lead to increased difficulty in maintaining, extending, or understanding the codebase \cite{chen2016detecting}.

Smells can be categorized into three main levels:

\begin{enumerate}
\item \textbf{Architectural Smells:} These are high-level issues that affect the overall structure and organization of the software system \cite{mo2019architecture} \cite{garcia2009toward}. Key manifestations include cyclic dependencies between modules or components, god components that attempt to handle too many responsibilities, and scattered functionality where related features are dispersed across unrelated modules.

\item \textbf{Structural Smells:} These relate to problems in the design of classes and modules \cite{gupta2022prioritizing, gupta2023severity, mo2019architecture, suryanarayana2014refactoring, lanza2007object}. The primary concerns involve issues in class and module design patterns, interface cohesion challenges, and complications in complexity and inheritance hierarchies.

\item \textbf{Code-Level Smells:} These represent the most granular issues within individual methods or small code blocks \cite{cardozo2023prevalence, foidl2022data, martin2009clean}. Common manifestations include long methods that have grown unwieldy, duplicate code segments that could be unified, and overly complex conditional logic that hampers readability and maintenance.

\end{enumerate}

The impact of code smells on software quality is significant and multifaceted. Architectural smells can lead to systems that are difficult to evolve and maintain, often resulting in increased development time and cost \cite{arcelli2023impact, mumtaz2021systematic}. Structural smells can make individual components harder to understand and modify, potentially introducing bugs during changes \cite{sas2022evolution, martini2018identifying}.  Code-level smells can decrease readability and increase the likelihood of errors in specific functions or methods \cite{chen2016detecting, boutaib2021code, palomba2013detecting}.

In the context of Python development, code smells take on additional significance due to the language's emphasis on readability and its "batteries included" philosophy. Python's dynamic nature and support for multiple programming paradigms can sometimes lead to unique code smells that may not be as prevalent in other languages.

\subsection{Existing Detection Tools}

The landscape of Python code analysis tools presents a diverse array of solutions, each with distinct capabilities and limitations as shown in Table \ref{tab:code_smell_tools}. Pylint, while widely adopted for style enforcement and error detection, primarily addresses surface-level concerns rather than deeper design issues \cite{pylint}. Similarly, Flake8, which integrates PyFlakes, pycodestyle, and McCabe complexity checker, offers valuable style checks and complexity measurements but lacks comprehensive smell detection capabilities \cite{flake8, pyflakes}.

SonarQube represents a more holistic approach to code analysis, though its Python-specific features remain less sophisticated compared to its robust Java support \cite{sonarqube}. This limitation is particularly evident in architectural smell detection, where specialized tools like Arcan demonstrate stronger capabilities, albeit primarily for Java codebases \cite{fontana2017arcan}.

Prospector attempts to bridge these gaps by integrating multiple analysis tools, yet struggles with detecting higher-level architectural smells and complex inheritance hierarchies \cite{prospector}. Radon, while effective in computing various code metrics such as cyclomatic complexity and maintainability index, focuses primarily on metrics rather than smell detection \cite{radon}.

These existing solutions generally excel at code-level analysis but demonstrate significant limitations in addressing higher-level architectural and structural concerns. The current tooling landscape reveals a clear need for more comprehensive solutions that can effectively address the full spectrum of code smells in Python projects, from architectural patterns to code-level issues. This gap is particularly notable in the detection of complex inheritance hierarchies, poor interface cohesion, and other sophisticated architectural patterns that impact code quality and maintainability.

\begin{table*}[t]
\centering
\caption{Comparison of Python Code Smell Detection Tools}
\label{tab:code_smell_tools}
\resizebox{\textwidth}{!}{%
\begin{tabular}{|l|c|c|c|c|c|c|}
\hline
\textbf{Tool} & \textbf{Type} & \textbf{Test-Specific} & \textbf{Customization} & \textbf{Architectural Smell} & \textbf{Structural Smell} & \textbf{Code Smell} \\
\hline
pytest-smell\cite{pytestsmell} & Free & Yes & Limited & No & No & Yes \\
Pylint\cite{pylint} & Free & No & Yes & No& No& Yes \\
Pyflakes\cite{pyflakes} & Free & No & Limited & No & Limited & Yes \\
Radon\cite{radon} & Free & No & Yes & No & No & Yes \\
Cohesion\cite{cohesion} & Free & No & Limited & No& Limited & No\\
Pyscent\cite{pyscent} & Free & No & Limited& No & No & Yes \\
PyNose\cite{pynose} & Free & Yes & Yes & No & No & Yes \\
good-smell\cite{goodsmell} & Free & No & Yes & No & No & Yes \\
SonarQube\cite{sonarqube} & Free/Paid & No & Yes & No & No & Yes \\
Prospector\cite{prospector} & Free & No & Yes & No& Limited & Yes \\
Flake8\cite{flake8} & Free & No & Yes & No & Limited & Yes \\
CodeScene\cite{codescene} & Paid & No & Yes & No & Limited & Yes \\
Arcan\cite{fontana2017arcan} & Paid & No & No & Yes & No & No \\
\textbf{PyExamine} & Free & No& \textbf{Yes}& \textbf{Yes} & \textbf{Yes} & \textbf{Yes} \\
\hline
\end{tabular}%
}

\end{table*}

These tools face four primary limitations:

\begin{enumerate}
\item Limited multi-level analysis, with most tools focusing primarily on code-level or simple structural issues
\item Insufficient consideration of Python-specific contexts and idioms
\item Restricted customizability for project-specific smell definitions
\item Inadequate explanatory guidance for detected issues
\end{enumerate}

CodeScene primarily focuses on behavioral code analysis using version control data to identify hotspots and development patterns. While it can highlight dependencies through co-change analysis \cite{codescene}, it doesn't specifically detect architectural smells in the same way that PyExamine does.

 As Garcia et al. \cite{garcia2009toward} note, architectural smells' high-level nature and need for broader context make automated detection particularly challenging.

Structural smell detection faces its own set of challenges. Despite Suryanarayana et al. \cite{suryanarayana2014refactoring} emphasizing their importance for software quality, many popular tools provide incomplete coverage in this area \cite{palomba2018scent}.

Recent research has attempted to address these limitations. Studies have explored various approaches for architectural smell detection \cite{fontana2017arcan, arcelli2023impact}, though primarily focusing on Java systems. In the Python ecosystem, development of specialized architectural smell detection capabilities remains limited compared to the comprehensive range of smells identified in literature \cite{vavrova2017does, giordano2023understanding}.

These gaps highlight the need for more sophisticated detection mechanisms, particularly for Python development. While various tools exist in the landscape, the fundamental challenge of comprehensive smell detection remains largely unaddressed \cite{mumtaz2021systematic, sas2022relation}.

PyExamine aims to bridge these gaps by offering a comprehensive, Python-specific solution that operates across multiple abstraction levels. By combining architectural, structural, and code-level analysis with deep understanding of Python's unique characteristics, it seeks to provide more complete and relevant code quality assessment for Python projects of all sizes. 

The tool emphasizes customizability and actionable insights, addressing the limitations identified in existing solutions while maintaining the accessibility and integration capabilities that developers expect from modern development tools.

\section{PyExamine Architecture}

\subsection{System Overview}

PyExamine is designed with a modular and extensible architecture to facilitate comprehensive code smell detection across multiple levels of abstraction. The system is composed of several key components that work together to analyze Python codebases and identify potential issues.

In total, PyExamine encompasses 49 distinct metrics distributed across the three analysis levels: 24 code-level metrics, 6 architectural metrics, and 19 structural metrics. 

These metrics are meticulously defined within the ``code\_quality\_config.YAML`` file, a crucial component of the comprehensive replication package available on Zenodo.  For detailed information and further clarification, consult the project's documentation.

This multi-layered approach enables a thorough evaluation of Python codebases, from high-level architectural considerations down to specific implementation details, providing a complete picture of code quality.
This implementation goes beyond simple metrics to consider:

\begin{itemize}
    \item Contextual complexity based on surrounding code
    \item Framework-specific anti-patterns
    \item Cross-module impact of architectural decisions
    \item Weighted severity based on multiple factors
\end{itemize}

\subsection*{Name Binding and Type Resolution Implementation in PyExamine}

The implementation of name binding and type resolution in PyExamine is more sophisticated than it may initially seem. Examining the code in \texttt{architectural\_smell\_detector.py}, we can observe a multi-layered approach to name resolution that handles both intra-module and inter-module relationships. 

The system employs a hierarchical resolution strategy that begins with local scope analysis and extends to package-level name binding. For instance, the method \texttt{resolve\_external\_dependencies(self)} demonstrates how PyExamine distinguishes between project modules, standard library imports, and third-party dependencies. The implementation carefully tracks module hierarchies and resolves relative imports by maintaining context about the package structure. 

Furthermore, the system manages Python's dynamic nature by analyzing runtime patterns and potential name rebinding scenarios.

\begin{figure}
    \centering
    \includegraphics[width=1\linewidth]{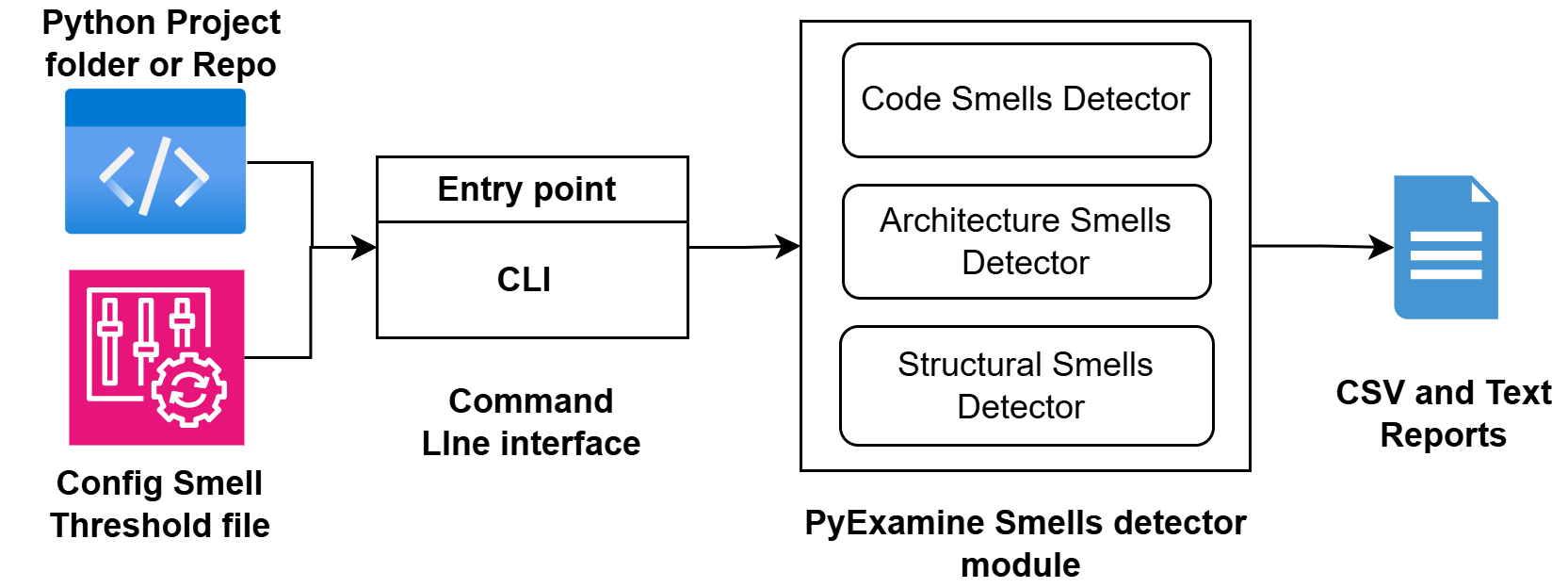}
    \caption{Main Component of PyExamine}
    \label{fig:PyExamine_architecture}
\end{figure}

As illustrated in Figure \ref{fig:PyExamine_architecture}, the main components of PyExamine are:

\begin{itemize}
    \item \textbf{Command Line Analysis Coordinator:} The central component that orchestrates the overall analysis process. It manages the parsing of Python files, delegates specific analysis tasks to the appropriate smell detection modules, and aggregates results.
    
    \item \textbf{Architectural Smell Detector:} Focuses on identifying high-level design issues within the project, such as cyclic dependencies and hub-like modules.
    
    \item \textbf{Structural Smell Detector:} Concentrates on evaluating class and module-level design patterns, with a particular emphasis on object-oriented principles and structural metrics
    
    \item \textbf{Code Smell Detector:} Operates at the most granular level, analyzing individual functions and code blocks for issues like long methods and duplicate code.
    
    \item \textbf{Configuration System:} Allows users to customize smell definitions, thresholds, and detection rules through a YAML-based configuration file.
    
    \item \textbf{Report Generator:} Compiles the results from all detectors into a comprehensive, easy-to-understand report.
\end{itemize}

The rationale behind this architecture is to provide a comprehensive yet modular approach to code smell detection. By separating concerns into distinct modules, PyExamine can offer in-depth analysis at multiple levels while remaining extensible and maintainable. The use of a central coordinator ensures efficient processing of large codebases, while the configuration system allows for fine-tuned analysis that can adapt to different project requirements and coding standards.

\subsection{Smell Detection Modules}

Code quality analysis can be divided into three primary modules: Architectural Analysis, Structural Analysis, and Code-Level Analysis. Each module serves a distinct purpose in evaluating different aspects of code quality \cite{mumtaz2021systematic}. The detection thresholds for various metrics have been carefully derived from established research and empirical studies \cite{fontana2015automatic, spadini2020investigating}, ensuring reliable and validated detection capabilities.

\subsubsection{Architectural Smell Detector}

The Architectural Analysis Module focuses on examining high-level design patterns and system architecture in Python projects \cite{garcia2009toward}.

It specifically analyzes inter-module relationships and system-wide patterns using graph-based algorithms \cite{fontana2017arcan}. The module's key capabilities include detecting cyclic dependencies with maximum length thresholds and identifying hub-like dependency patterns through ratio analysis \cite{arcelli2023impact}. 

It also evaluates unstable dependencies using threshold-based detection and monitors API usage patterns for improper implementation \cite{martini2018identifying}. 

Furthermore, this module examines module interaction patterns, identifying redundant abstractions through similarity analysis and detecting god objects by analyzing function counts \cite{sas2022evolution}.

The metric computation in PyExamine is built on an Abstract Syntax Tree (AST) analysis system that goes well beyond simple syntax parsing. The architectural metrics are computed through a combination of:

\begin{itemize}
    \item Direct AST analysis for structural information
    \item Dependency graph analysis for architectural relationships
    \item Cross-reference analysis for coupling metrics
    \item Context-aware metric aggregation
\end{itemize}

The system builds a comprehensive dependency graph that captures both explicit and implicit relationships between modules. This enables the detection of many architectural smells used in our tool.

\subsubsection{Structural Smell Detector}

The Structural Analysis Module concentrates on evaluating class and module-level design patterns, with a particular emphasis on object-oriented principles and structural metrics \cite{chidamber1994metrics}. 

It employs several complexity metrics, including Weighted Methods Per Class (WMPC1, WMPC2), Cyclomatic Complexity, and Size and Length Metrics (SIZE2, LOC) \cite{lanza2007object}. 

The module also assesses coupling and cohesion through metrics such as Lack of Cohesion of Methods (LCOM), Coupling Between Objects (CBO), and Message Passing Coupling (MPC) \cite{sahraoui2000can}.

Inheritance and dependencies are evaluated using metrics like Depth of Inheritance Tree (DIT), Number of Children (NOC), and Fan-in/Fan-out Analysis \cite{chidamber1994metrics}. 

Additionally, the module examines code structure elements including maximum nesting depth, branch complexity, and file length thresholds \cite{fontana2015automatic}.

\subsubsection{Code Smell Detector}

The Code-Level Analysis Module performs detailed analysis of function and method-level code quality. It identifies specific code smells and maintenance issues across several categories \cite{fowler1999refactoring}.

In terms of code organization, it analyzes long methods through line count analysis, large classes by method count, parameter list complexity, and primitive type usage patterns \cite{martin2009clean}. 

The module also addresses code duplication and maintenance by identifying duplicate code segments with minimum lines threshold, dead code, speculative generality, and divergent change patterns \cite{boutaib2021code}.

Code complexity assessment includes analysis of complex conditional structures, message chains, feature envy detection, and data clump identification \cite{palomba2013detecting}. The module also evaluates documentation and comments through ratio analysis, large comment block detection, and documentation quality metrics \cite{vavrova2017does}.

\subsection{Configuration System}

The Configuration System is a critical component of PyExamine that provides flexibility and customizability to the tool's smell detection process. It allows users to tailor the behavior of the smell detectors to their specific project needs, coding standards, and quality requirements.

Key features of the Configuration System include:

\begin{itemize}
    \item \textbf{YAML-based Configuration:} Uses a human-readable YAML format for easy editing and version control.
    \item \textbf{Customizable Thresholds:} Allows users to set project-specific thresholds for various metrics used in smell detection.
    \item \textbf{Smell Prioritization:} Enables users to assign different weights or priorities to different types of smells.
    \item \textbf{Custom Smell Definitions:} Allows users to define project-specific smells based on their unique requirements.
\end{itemize}

The Configuration System interacts closely with all smell detection modules, allowing them to dynamically adjust their behavior based on the user-defined settings. This high degree of customization makes PyExamine adaptable to a wide range of Python projects and development practices.

\subsection{Replication Package}
To ensure the reproducibility of our results and to facilitate further research in this area, we have prepared a comprehensive replication package for \texttt{PyExamine}. This package includes all the necessary components to recreate our experiments and verify our findings.

The replication package is available through multiple channels:

\textbf{Documentation:} Detailed documentation of \texttt{PyExamine}, including installation instructions, usage guidelines, and API references, is available at\footnote{\href{https://python-smell.readthedocs.io/en/}{https://python-smell.readthedocs.io/en/}}.

\textbf{Source Code:} The complete source code of \texttt{PyExamine} \footnote{\url{https://github.com/KarthikShivasankar/python_smells_detector}}.

\textbf{Data and Results:} To ensure long-term availability and citability, we have archived everything, including the analyzed Python projects, on Zenodo\footnote{\url{https://doi.org/10.5281/zenodo.14060772}}. 

By providing this comprehensive replication package, we aim to contribute to the transparency and reproducibility of software engineering research.

\section{Evaluation}

\subsection{Methodology}

To assess the effectiveness and performance of PyExamine, we conducted a comprehensive evaluation using a diverse set of Python projects. Our methodology was designed to test the tool's capabilities across various dimensions, including detection accuracy, performance, and usability. The evaluation process consisted of several key components:

\begin{enumerate}
    \item \textbf{Dataset Selection:} The analysis of 183 open-source Python projects from GitHub was used as a broad study of code smell patterns and prevalence in the Python ecosystem. This large-scale analysis provided insights into common issues across different types and sizes of Python projects, but it was not used to directly evaluate PyExamine's performance.
    
For the actual evaluation of PyExamine, the paper focused on a smaller set of 7 diverse Python repositories, as detailed in Table \ref{tab:survey_repo_github}. These repositories were used to assess the tool's detection accuracy, compare it against expert evaluations, and gather user feedback.

    \item \textbf{Tool Configuration:} For each project in our evaluation, we utilized PyExamine with both default configurations and custom configurations tailored to the specific requirements of each codebase. The configuration was implemented using YAML files, allowing for fine-tuned adjustment of smell detection thresholds and criteria. The detection thresholds for various metrics have been carefully derived from established research and empirical studies \cite{fontana2015automatic, spadini2020investigating}, ensuring reliable and validated detection capabilities.
An example of a custom configuration for one of the projects is as follows:

\begin{verbatim}
code_smells:
LONG_METHOD_LINES:
value: 45
LARGE_CLASS_METHODS:
value: 15
\end{verbatim}

This configuration approach allowed us to adapt PyExamine's detection criteria to the specific coding standards and architectural patterns of each project under evaluation. By adjusting these thresholds, we could account for variations in project size, complexity, and domain-specific requirements, ensuring a fair and contextually appropriate assessment across our diverse set of repositories.

    \item \textbf{User Study:} We conducted a targeted evaluation of PyExamine involving seven diverse Python projects and their main contributors. The study encompassed a range of domains, from scientific computing libraries to enterprise solutions, ensuring a comprehensive assessment of the tool's capabilities across different contexts.
    
The main contributor for each selected project was invited to participate, providing a pool of developers with Python experience ranging from 4 to 13 years (mean: 8.43 years). This diversity in experience levels allowed for a robust evaluation of PyExamine's effectiveness across different skill sets.

To establish a reliable ground truth, participants were tasked with manually identifying code smells in their own projects. This approach leveraged the developers' intimate knowledge of their codebases, creating a high-quality dataset for comparison. PyExamine was then run on each project, and its results were meticulously compared against the developer-provided ground truth.

Using this comparison, we calculated key classification metrics, including false positives, recall, F1-score, and F2-score. These metrics provided a quantitative measure of PyExamine's performance in detecting various types of code smells. 

Additionally, participants offered valuable feedback on PyExamine's accuracy, usability, relevance of detected smells, and clarity of generated reports.
This comprehensive approach allowed us to evaluate PyExamine's performance against expert knowledge while gathering insights from experienced Python developers across different domains. The combination of quantitative metrics and qualitative feedback provided a holistic view of PyExamine's effectiveness as a code smell detection tool.

\end{enumerate}

\subsection{Detection Accuracy}

The assessment of PyExamine involved a comprehensive evaluation across seven diverse Python repositories. From these repositories, we randomly selected a total of 94 code snippets suspected of containing smells. These snippets represented 28 unique code smell types, providing a broad spectrum for evaluation.

Each of these 94 code snippets was independently assessed by the project's main developer. This approach allowed us to compare PyExamine's automated detection capabilities against expert human judgment, offering crucial insights into the tool's reliability and effectiveness. 

Each project's main developer analyzed only their own project's files, not the entire set of 94 files. The distribution of smells was not uniform across projects: some smells appeared frequently, while others were rare. The prevalence of specific smells largely depended on each project's context and requirements.

This extensive evaluation process provided valuable data on PyExamine's detection capabilities across various smell types and project contexts. By comparing the tool's results with developer assessments, we gained a clear picture of PyExamine's accuracy and its alignment with expert opinion in identifying code quality issues.

As shown in Table \ref{tab:survey_repo_github}, our evaluation utilized repositories from developers with varying levels of Python expertise. The repositories included scientific computing (pseudo-hamiltonian-neural-networks, 6 years), system tools (sponge with 9 years, rewts with 12 years), domain-specific applications (melodi with 8 years, rttarr with 4 years), enterprise solutions (cyberrisk with 13 years), and private development projects (inomotifin with 7 years). This diverse experience profile, ranging from 4 to 13 years with a mean experience of 8.43 years, adds significant credibility to our evaluation.

\begin{table}[h]
\centering
\caption{Python Experience and GitHub Repositories of Survey Respondents}
\label{tab:survey_repo_github}
\begin{tabular}{cl}
\hline
\textbf{Python Experience (Years)} & \textbf{Repository Name} \\
\hline
9  & \href{https://github.com/ladislav-hovan/sponge}{sponge} \\
\hline
8  & \href{https://github.com/ejhusom/MELODI/tree/pyjoules-integration}{melodi} \\
\hline
6  & \href{https://github.com/SINTEF/pseudo-hamiltonian-neural-networks}{pseudo-hamiltonian-neural-networks} \\
\hline
12 & \href{https://github.com/SINTEF/rewts}{rewts} \\
\hline
4  & \href{https://github.com/secureIT-project/RTT_for_APR}{rttarr} \\
\hline
7  & inomotifin (Private repo) \\
\hline
13 & \href{https://github.com/mnemonic-no/cyberrisk}{cyberrisk} (Part of the repo) \\
\hline
\end{tabular}
\end{table}

\begin{table}[h]
\centering
\caption{Classification Metrics}
\label{tab:classification_metric_analysis}
\begin{tabular}{lr}
\hline
\textbf{Metric} & \textbf{Value} \\
\hline
Recall & 0.872 \\
F1-score & 0.932 \\
F2-score & 0.895 \\
\hline
\end{tabular}
\end{table}

PyExamine demonstrated strong detection performance across key metrics, as detailed in Table \ref{tab:classification_metric_analysis}. The tool achieved a recall rate of 0.872, successfully identifying 87.2\% of actual code smells, aligning remarkably well with our evaluator agreement rate of 87.23\%. The exceptional F1-score of 0.932 indicates balanced performance with minimal false positives and negatives, while the strong F2-score of 0.895 confirms the tool's effectiveness in real-world scenarios where missing code smells is considered more costly than false positives.

The evaluation's reliability is substantiated by a high evaluator agreement rate of 87.23\%, with a 95\% confidence interval ranging from 80.49\% to 93.98\%. This strong consensus among evaluators validates the tool's detection patterns and methodology robustness. The narrow confidence interval particularly supports the reliability of our ground truth data and evaluation approach.

Our evaluation framework compared PyExamine's automated detection results against manual code reviews, serving as ground truth. This comparison assessed both traditional code smells and architectural issues, providing comprehensive validation of the tool's detection algorithms. The classification metrics presented in Table \ref{tab:classification_metric_analysis} demonstrate the tool's robust performance across different detection scenarios.

The combination of high performance metrics from Table \ref{tab:classification_metric_analysis} and strong evaluator agreement demonstrates PyExamine's capability to replicate expert-level code smell detection. The tool's balanced performance makes it particularly valuable for both automated code review processes and as a supporting tool for manual reviews. This reliability, coupled with comprehensive coverage across different application domains as evidenced in Table \ref{tab:survey_repo_github}, establishes PyExamine as an effective solution for automated code quality assessment.

The evaluation results validate PyExamine's practical value in software development workflows. The tool's ability to maintain consistent performance across diverse repositories with varying developer experience levels (as shown in Table \ref{tab:survey_repo_github}) and align with expert judgment makes it a reliable instrument for continuous code quality monitoring. The high F1 and F2 scores, combined with strong evaluator agreement, suggest that PyExamine successfully bridges the gap between automated detection and human expertise in code smell identification.

This comprehensive evaluation proves PyExamine's effectiveness as a code smell detection tool, supported by robust metrics, diverse repository assessment, and strong evaluator agreement. The tool's demonstrated capabilities in detecting both simple and complex code smells, combined with its alignment with expert assessment, make it a valuable asset for modern software development practices.

\begin{table}[h]
\centering
\caption{Smell Type Analysis}
\label{tab:smell_type_analysis}
\begin{tabular}{lrrr}
\hline
\textbf{Smell Type} & \textbf{Count} & \textbf{Agreement Rate (\%)} & \textbf{Proportion (\%)} \\
\hline
Architectural Smell & 31 & 80.65 & 32.98 \\
Code Smell & 35 & 91.43 & 37.23 \\
Structural Smell & 28 & 89.29 & 29.79 \\
\hline
\end{tabular}
\end{table}

\begin{table}[h]
\centering
\caption{Reliability Metrics (Per Smell Type)}
\label{tab:reliability_metric_analysis}
\begin{tabular}{lrrr}
\hline
\textbf{Smell Type} & \textbf{Recall} & \textbf{F1-score} & \textbf{F2-score} \\
\hline
Code Smell & 0.914 & 0.955 & 0.930 \\
Structural Smell & 0.893 & 0.943 & 0.912 \\
Architectural Smell & 0.806 & 0.893 & 0.839 \\
\hline
\end{tabular}
\end{table}

The analysis of PyExamine's detection capabilities can be structured around three distinct sub research questions, each focusing on a specific type of code smell as evidenced in Table \ref{tab:smell_type_analysis} and Table \ref{tab:reliability_metric_analysis}.

\subsubsection{RQ 1: How accurate is PyExamine in detecting code smells??}

Our analysis of code smell detection reveals PyExamine's exceptional performance in identifying traditional code-level issues. As shown in Table \ref{tab:reliability_metric_analysis}, the tool achieved good classification metrics among all smell types, with a recall of 0.914, F1-score of 0.955, and F2-score of 0.930. Table \ref{tab:smell_type_analysis} further supports this finding, showing a 91.43\% agreement rate among evaluators and representing 37.23\% of all detected smells (35 instances). These robust metrics indicate PyExamine's particular strength in identifying common programming antipatterns, making it highly reliable for routine code quality assessment tasks.

\subsubsection{RQ 2: How accurate is PyExamine in detecting structural smells?}

The evaluation of structural smell detection demonstrates PyExamine's strong capabilities in identifying code organization and component relationship issues. Table \ref{tab:reliability_metric_analysis} shows good performance metrics with a recall of 0.893, F1-score of 0.943, and F2-score of 0.912. According to Table \ref{tab:smell_type_analysis}, structural smells comprised 29.79\% of total detections (28 instances) with an agreement rate of 89.29\%. While these metrics are slightly lower than those for code smells, they nonetheless indicate robust detection capabilities for structural issues, suggesting PyExamine effectively identifies problems related to code organization and component interactions.

\subsubsection{RQ 3: How accurate is PyExamine in detecting architectural smells??}

The analysis of architectural smell detection reveals PyExamine's capability to identify complex, system-level issues, albeit with lower metrics compared to other smell types. As evidenced in Table \ref{tab:reliability_metric_analysis}, architectural smell detection achieved a recall of 0.806, F1-score of 0.893, and F2-score of 0.839. Table \ref{tab:smell_type_analysis} shows that architectural smells constituted 32.98\% of all detections (31 instances) with an 80.65\% agreement rate. While these metrics are lower than those for code and structural smells, they represent satisfactory performance given the inherent complexity of architectural pattern detection.

The comparative analysis across all three research questions reveals a clear performance gradient, with traditional code smells showing the highest detection accuracy, followed by structural smells, and then architectural smells. This pattern is consistent across both detection metrics (Table \ref{tab:reliability_metric_analysis}) and evaluator agreement rates (Table \ref{tab:smell_type_analysis}). The distribution of smell types shows a relatively balanced representation, with code smells slightly more prevalent at 37.23\%, followed by architectural smells at 32.98\%, and structural smells at 29.79\%.

This comprehensive evaluation demonstrates PyExamine's versatility in detecting various types of code smells, with particularly strong performance in identifying code-level and structural issues. The tool maintains acceptable accuracy even for complex architectural patterns, though with expectedly lower performance metrics. The correlation between agreement rates and detection metrics across smell types suggests that PyExamine's detection algorithms align well with expert judgment, particularly for more straightforward code-level issues, while maintaining reliable performance for more complex structural and architectural patterns.

\subsubsection{Tools usefulness and qualitative assessment }

\begin{table}[h]
\centering
\caption{Survey Results Analysis (1-4 Likert Scale)}
\label{tab:survey_results}
\begin{tabular}{lcccc}
\hline
Metric & Usefulness & Code  & Architectural  & Structural  \\
\hline
Mean & 3.57 & 3.86 & 3.43 & 3.71 \\
Std Dev & 0.53 & 0.38 & 0.53 & 0.49 \\
Min & 3.00 & 3.00 & 3.00 & 3.00 \\
Max & 4.00 & 4.00 & 4.00 & 4.00 \\
\hline
\end{tabular}

\end{table}

The survey results in Table \ref{tab:survey_results} demonstrate PyExamine's effectiveness across multiple dimensions, based on anonymous feedback from the 7 developers listed in Table \ref{tab:survey_repo_github}. The tool received strong overall ratings with a mean of 3.57 (SD=0.53), ranging from 3.0 to 4.0.

Code smell detection emerged as the strongest feature (mean=3.86, SD=0.38), aligning with the metrics shown in Table \ref{tab:reliability_metric_analysis}. Structural smell detection followed closely (mean=3.71, SD=0.49), corresponding well with its F1-score of 0.943. Architectural smell detection, while slightly lower (mean=3.43, SD=0.53), maintained satisfactory performance.

The consistent maximum rating of 4.0 across all categories and minimum of 3.0 validates PyExamine's effectiveness as a comprehensive code smell detection tool, with particular strength in code-level analysis while maintaining reliable performance for structural and architectural patterns.

PyExamine demonstrated high recall accuracy across all categories, with particularly strong performance in detecting code-level smells. The slightly lower recall for architectural smells can be attributed to the inherent complexity of detecting high-level design issues, which sometimes require contextual understanding that is challenging to fully automate.

In our  analysis, we found that the majority of false positives were due to project-specific conventions or intentional design decisions that deviated from general best practices. This highlights the importance of PyExamine's customization capabilities, which allow users to tailor the tool to their project's specific needs.

\subsection{RQ 4: What is the prevalence and distribution of different smell types in Python projects?}

\begin{table}[h]
\centering
\caption{Summary Statistics of Code Smell Analysis}
\label{tab:summary_stats}
\begin{tabular}{lr}
\hline
\textbf{Metric} & \textbf{Value} \\
\hline
Total Smells Detected & 1,151,059 \\
Total Files Affected & 33,765 \\
Unique Smell Names & 37 \\
Total Projects & 183 \\
Total Modules & 66,352 \\
\hline
\end{tabular}

\end{table}

The analysis of code smell prevalence in Python projects revealed significant findings, as shown in Table \ref{tab:summary_stats}. Our study detected 1,151,059 code smells across 183 Python projects, with 33,765 out of 66,352 modules (50.9\%) containing at least one smell.

The investigation identified 37 unique smell types across architectural, structural, and code smell categories. 

The comprehensive dataset, spanning a substantial number of projects and modules, provides robust evidence of code smell distribution in Python projects. These findings emphasize the pervasive nature of code quality issues and the need for systematic detection and remediation approaches.

\subsubsection{Most Common Code Smells Analysis}

\begin{table}[h]
\centering
\caption{ Most Common Code Smells}
\label{tab:top_10_smells}
\begin{tabular}{lrr}
\hline
\textbf{Smell Type} & \textbf{Count} & \textbf{Percentage} \\
\hline
Feature Envy & 42,368 & 3.68\% \\
Potential Shotgun Surgery & 22,745 & 1.98\% \\
Too Many Branches & 21,691 & 1.88\% \\
Potential Improper API Usage & 17,377 & 1.51\% \\
Scattered Functionality & 16,470 & 1.43\% \\
Unstable Dependency & 12,611 & 1.10\% \\
Long Method & 12,033 & 1.05\% \\
Temporary Field & 10,893 & 0.95\% \\
Potential Redundant Abstractions & 7,986 & 0.69\% \\
\hline
\end{tabular}

\end{table}

Table \ref{tab:top_10_smells} reveals a feature Envy emerges as the most common smell with 3.68\% (42,368 instances), followed by Potential Shotgun Surgery at 1.98\% (22,745 instances). 

The distribution of remaining smells shows a gradual decline in frequency, with Too Many Branches (1.88\%), Potential Improper API Usage (1.51\%), and Scattered Functionality (1.43\%) representing significant but less prevalent issues. The presence of both implementation-level smells (Long Method at 1.05\%) and design-level issues (Unstable Dependency at 1.10\%) in the Table  \ref{tab:top_10_smells} suggests that code quality challenges exist across different abstraction levels.

\subsubsection{Categorical Distribution of Code Smells}

\begin{table}[h]
\centering
\caption{Distribution of Smell Types by Category}
\label{tab:smell_distribution}
\begin{tabular}{llrr}
\hline
\textbf{Category} & \textbf{Smell Type} & \textbf{Count} & \textbf{\%} \\
\hline
\multirow{5}{*}{Structural} & Too Many Branches & 21,691 & 32.91\% \\
& High Lines of Code (LOC) & 7,921 & 12.02\% \\
& High Response for a Class (RFC) & 7,899 & 11.99\% \\
& High Cyclomatic Complexity & 6,318 & 9.59\% \\
& High Number of Methods (NOM) & 6,221 & 9.44\% \\
\hline
\multirow{5}{*}{Code} & Feature Envy & 42,368 & 4.13\% \\
& Potential Shotgun Surgery & 22,745 & 2.22\% \\
& Long Method & 12,033 & 1.17\% \\
& Temporary Field & 10,893 & 1.06\% \\
\hline
\multirow{5}{*}{Architectural} & Potential Improper API Usage & 17,377 & 28.92\% \\
& Scattered Functionality & 16,470 & 27.41\% \\
& Unstable Dependency & 12,611 & 20.99\% \\
& Potential Redundant Abstractions & 7,986 & 13.29\% \\
& God Object & 2,476 & 4.12\% \\
\hline
\end{tabular}

\end{table}

Table \ref{tab:smell_distribution} provides a detailed breakdown of smell distributions across structural, code, and architectural categories, offering deeper insights into the nature of code quality issues in Python projects:

In the Structural category, Too Many Branches leads with 32.91\% (21,691 instances), followed by High Lines of Code (12.02\%) and High Response for a Class (11.99\%). This distribution suggests that complexity management and code organization are significant challenges in Python development. The relatively even distribution among other structural metrics indicates that these issues often occur in combination.

The Code smell category is  presented by  Feature Envy (4.13\%) and Potential Shotgun Surgery (2.22\%). This stark imbalance suggests that data organization and method-data relationships are particularly challenging aspects of Python development. The lower percentages of Long Method (1.17\%) and Temporary Field (1.06\%) indicate that these traditional code smells are less prevalent but still significant.

Architectural smells show a more balanced distribution, with Potential Improper API Usage (28.92\%) and Scattered Functionality (27.41\%) being the most common issues. The significant presence of Unstable Dependency (20.99\%) and Potential Redundant Abstractions (13.29\%) suggests that architectural design challenges are widespread in Python projects. The lower occurrence of God Object (4.12\%) might indicate better awareness and avoidance of this well-known anti-pattern.

\begin{table}[htbp]
\centering
\small
\caption{\footnotesize Distribution of Code Smells Across Python Frameworks}
\begin{tabular}{@{}l|r|rrr@{}}
\hline
\footnotesize\textbf{Framework} & \footnotesize\textbf{Total Smells} & \multicolumn{3}{c@{}}{\footnotesize\textbf{Distribution (\%)}} \\
& & \footnotesize\textbf{Arch.} & \footnotesize\textbf{Code} & \footnotesize\textbf{Struct.} \\
\hline
\footnotesize Keras & \footnotesize 20,458 & \footnotesize 7.7 & \footnotesize 86.8 & \footnotesize 5.5 \\
\footnotesize NumPy & \footnotesize 16,368 & \footnotesize 6.1 & \footnotesize 85.5 & \footnotesize 8.4 \\
\footnotesize Scikit-learn & \footnotesize 11,083 & \footnotesize 10.0 & \footnotesize 79.8 & \footnotesize 10.2 \\
\footnotesize FastAPI & \footnotesize 1,758 & \footnotesize 51.8 & \footnotesize 38.8 & \footnotesize 9.4 \\
\footnotesize Flask & \footnotesize 431 & \footnotesize 36.9 & \footnotesize 45.5 & \footnotesize 17.6 \\
\footnotesize Requests & \footnotesize 148 & \footnotesize 17.6 & \footnotesize 34.5 & \footnotesize 47.9 \\
\hline
\multicolumn{5}{@{}l@{}}{\scriptsize Note: Arch.=Architectural, Struct.=Structural} \\
\end{tabular}
\label{table:smell_distribution}
\end{table}

\begin{table}[htbp]
\centering
\small
\caption{\footnotesize Most Prevalent Code Smells by Framework}
\begin{tabular}{@{}l|l@{}}
\hline
\footnotesize\textbf{Framework} & \footnotesize\textbf{Most Prevalent Smells} \\
\hline
\footnotesize Keras & \footnotesize SF(2.3\%), FE(2.4\%), TMB(1.7\%) \\
\footnotesize NumPy & \footnotesize SF(3.5\%), FE(4.8\%), SS(5.3\%) \\
\footnotesize Scikit-learn & \footnotesize UD(3.6\%), FE(22.2\%), LM(5.6\%) \\
\footnotesize FastAPI & \footnotesize RA(25.5\%), FE(14.7\%), LM(17.4\%) \\
\footnotesize Flask & \footnotesize CD(19.0\%), FE(31.3\%), TMB(6.5\%) \\
\footnotesize Requests & \footnotesize UD(4.7\%), FE(13.5\%), TMB(10.8\%) \\
\hline
\multicolumn{2}{@{}l@{}}{\scriptsize Smell abbreviations: SF=Scattered Functionality, FE=Feature Envy,} \\
\multicolumn{2}{@{}l@{}}{\scriptsize TMB=Too Many Branches, SS=Shotgun Surgery, UD=Unstable Dependency,} \\
\multicolumn{2}{@{}l@{}}{\scriptsize RA=Redundant Abstraction, LM=Long Method, CD=Cyclic Dependency} \\
\end{tabular}
\label{table:prevalent_smells}
\end{table}

The analysis of code smell distribution across Python frameworks reveals distinct patterns, as detailed in Table \ref{table:smell_distribution}. Machine learning and scientific libraries (Keras, NumPy, Scikit-learn) show high proportions of code-level smells (79.8-86.8\%), while web frameworks like FastAPI and Flask demonstrate more balanced distributions with higher architectural smell percentages (51.8\% and 36.9\%). Notably, the Requests library shows the highest proportion of structural smells (47.9\%).

Table \ref{table:prevalent_smells} highlights the most common smell types across frameworks. Feature Envy (FE) appears consistently throughout. Scientific libraries share similar patterns, with Scattered Functionality (SF) being prominent, while web frameworks show higher instances of architectural smells like Cyclic Dependency (CD) and Redundant Abstraction (RA). The Requests library notably struggles with Unstable Dependencies (UD) and Too Many Branches (TMB).

These patterns reflect the distinct challenges faced by different types of Python frameworks, suggesting the need for domain-specific code quality strategies.

This categorical analysis demonstrates that while certain smells dominate their respective categories, the nature and distribution of code quality issues vary significantly across different aspects of software design and implementation. The findings suggest that comprehensive code quality improvement strategies should address issues at all levels, with particular attention to data organization and structural complexity management in Python projects.

\section{Discussion}

The comprehensive evaluation of PyExamine has yielded significant insights into its effectiveness as a code smell detection tool. Our analysis demonstrates exceptional performance across multiple dimensions, with the tool achieving recall accuracy rates: 91.4\% for code-level smells, 89.3\% for structural smells, and 80.6\% for architectural smells. This hierarchical performance pattern naturally aligns with the increasing complexity of detection at different abstraction levels.

\subsection{User Study Insights}

User study involving seven experienced developers provided strong validation for PyExamine's practical value. Developers consistently expressed high satisfaction across all aspects of the tool's functionality, with overall usefulness receiving a robust mean rating of 3.57 out of 4.0. Code smell detection features stood out as particularly effective, garnering the highest rating of 3.86, while structural smell detection also received positive feedback with a mean rating of 3.71. Even the more complex architectural detection capabilities achieved a respectable 3.43, demonstrating broad user confidence in PyExamine's utility across all detection levels.

Qualitative feedback from the study revealed several key insights into PyExamine's strengths and areas for potential improvement. Developers particularly appreciated the tool's ability to identify straightforward code-level issues such as long methods, complex conditionals, and duplicate code \cite{martin2009clean}, which provided immediate, actionable insights for improving code quality. The detection of structural smells like poor interface cohesion and improper inheritance hierarchies \cite{chidamber1994metrics} was highlighted as an educational feature, helping developers enhance their object-oriented design skills.

However, users encountered some challenges with higher-level architectural smells \cite{arcelli2023impact}. Concepts such as unstable dependencies, scattered functionality, and cyclic dependencies were sometimes difficult to fully grasp and address, especially in larger codebases \cite{mumtaz2021systematic}. Several participants noted that while PyExamine excelled at identifying potential issues, interpreting the relevance of certain smells, particularly architectural ones, often required deeper project context \cite{martini2018identifying}. This was especially true for smells like feature envy and potential shotgun surgery \cite{fowler1999refactoring}.

While developers appreciated PyExamine's customizability, some found the initial configuration process for project-specific rules to be complex \cite{fontana2015automatic}, suggesting a need for more comprehensive documentation or preset configurations for common project types \cite{sas2022evolution}.

These insights underscore PyExamine's strength in providing immediate, actionable feedback on code quality while also revealing areas for potential improvement, particularly in guiding users through the interpretation and resolution of more complex architectural smells \cite{suryanarayana2014refactoring}. The study highlights the tool's value as both a practical development aid and a learning tool for advancing software design skills, while also indicating room for enhancing user guidance on more abstract code quality concepts \cite{vavrova2017does}.

\subsection{Tool Strengths}
PyExamine distinguishes itself from existing code smell detection tools through several key strengths. Its comprehensive multi-level analysis capability provides a holistic view of code quality by detecting smells at architectural, structural, and code levels \cite{mumtaz2021systematic}. This multi-tiered approach enables developers to address issues across various abstraction levels, from high-level design problems to low-level implementation details \cite{garcia2009toward}.

The tool's Python-specific detection mechanisms set it apart from generic code analysis tools. By understanding Python-specific idioms, best practices, and common pitfalls \cite{vavrova2017does},, PyExamine delivers more relevant and accurate smell detection results. The YAML-based configuration system offers exceptional customizability, allowing users to fine-tune smell definitions and thresholds according to their project's specific needs, making it adaptable to various coding standards and project requirements.

Beyond mere detection, PyExamine provides actionable insights by offering context and suggestions for improvement \cite{fowler1999refactoring}. This feature helps developers understand why certain code patterns are problematic and guides them in effective refactoring \cite{suryanarayana2014refactoring}. Furthermore, the tool's design facilitates seamless integration into existing development workflows, including continuous integration pipelines and code review processes.

\subsection{Limitations}
Despite its robust capabilities, PyExamine faces few limitations. While the tool maintains high accuracy, it can produce false positives, particularly for architectural smells that require deeper contextual understanding \cite{arcelli2023impact}. This issue becomes more pronounced when the tool isn't properly configured for a specific project's conventions.

The tool's high degree of customizability, while advantageous, introduces configuration complexity \cite{boutaib2021code}. Users must invest significant time in understanding and configuring the tool to maximize its benefits, which may challenge less experienced developers. As a static analysis tool, PyExamine cannot detect issues that only manifest at runtime, a limitation particularly relevant given Python's dynamic nature.

Python's rapid evolution necessitates frequent updates to keep PyExamine current with the latest language features and best practices \cite{vavrova2017does}. Additionally, the tool's current focus on analyzing the present state of codebases means it lacks insights into how code smells evolve over time or correlate with project history \cite{sas2022evolution}.

\subsection{Future Enhancements}

Our evaluation and user feedback have revealed several promising directions for PyExamine's future development. Developing plugins for popular IDEs would enable real-time smell detection and suggestions during code writing \cite{boutaib2021code}.

Implementing historical trend analysis capabilities would provide valuable insights into project health trends and refactoring effectiveness \cite{sas2022relation}. The tool's smell catalog could be continuously expanded, particularly focusing on Python-specific anti-patterns and emerging best practices \cite{martini2018identifying}. Enhanced visualization features, including sophisticated dependency graphs and heat maps, would improve the communication of complex code quality issues \cite{fontana2017arcan}.

Advanced capabilities could include automated refactoring suggestions and implementations to address identified issues  \cite{palomba2013detecting}. Performance optimization remains an ongoing priority, particularly for large codebases, through techniques like incremental analysis and more efficient algorithms. The tool's scope could be extended to analyze entire Python ecosystems, including dependencies and package interactions \cite{mumtaz2021systematic}, while seamless integration with CI/CD pipelines could be enhanced through customizable quality gates.

These future enhancements aim to evolve PyExamine into an even more powerful tool for maintaining high-quality Python codebases. By addressing current limitations and expanding its capabilities, PyExamine can continue to develop alongside Python's evolution and meet the growing needs of its developer community \cite{zakeri2023systematic}.

\section{Threats to Validity}

\subsection{Internal Validity}

Our study's internal validity considers potential biases in methodology. While covering 183 Python projects, dataset selection may contain inherent biases. Expert evaluations for ground truth, though showing 87.23\% agreement rate, could introduce subjectivity. The developer experience range (4-13 years, mean 8.43) may not fully represent all expertise levels.

\subsection{External Validity}

Generalizability limitations include potential gaps in domain coverage despite diverse project selection across scientific computing, system tools, and enterprise solutions. The user study's scope (seven developers) and evaluation coverage (28 smell types across 94 evaluations) may not fully represent the broader Python ecosystem.

\subsection{Construct Validity}
Performance metrics (recall, F1-score, F2-score) provide comprehensive measurement but may not capture all detection quality aspects. The categorization of smells (architectural: 32.98\%, code: 37.23\%, structural: 29.79\%) could influence result interpretation, despite high detection rates across categories.
\subsection{Reliability}
Our findings' reliability is supported by high evaluator agreement (87.23\%, CI: 80.49-93.98\%), multiple performance metrics, and positive user feedback across different aspects (usefulness: 3.57, code smell detection: 3.86, structural: 3.71, architectural: 3.43).

\section{Conclusion}

PyExamine is a comprehensive multi-level code smell detection tool designed for Python projects. It identifies potential issues at architectural, structural, and code levels, offering a holistic approach to code quality assessment.

The evaluation demonstrates PyExamine's effectiveness in detecting code smells with high recall accuracy. User study evaluation illustrate the tool's potential to improve code quality and maintainability.

It encourages better design practices by providing comprehensive code quality insights. The tool facilitates early detection of potential issues, allowing developers to address problems before they become deeply embedded in the project. 

 Additionally, it serves as an educational tool, helping developers learn and apply best practices in Python development, thereby enhancing overall code quality and software design skills.

\bibliographystyle{IEEEtran}
\bibliography{sample-base}

\end{document}